\documentclass[10pt,twocolumn,letterpaper,dvipsnames]{article} 

\usepackage[pagenumbers]{cvpr} 

\usepackage{graphicx}
\usepackage{amsmath}
\usepackage{amssymb}
\usepackage{booktabs}
\usepackage{siunitx}
\usepackage{sidecap} 
\usepackage[accsupp]{axessibility}

\usepackage[style=ieee, backend=bibtex, defernumbers=false]{biblatex}
\usepackage{dblfloatfix} 

\usepackage[breaklinks,colorlinks]{hyperref}

\usepackage[capitalize]{cleveref}
\crefname{section}{Sec.}{Secs.}
\Crefname{section}{Sec.}{Sects.}
\Crefname{table}{Table}{Tables}
\crefname{table}{Table}{Tabs.}

\usepackage{xcolor} 

\usepackage[draft]{pdfcomment} 
\usepackage[hyperfirst=false,acronym,sort=none,shortcuts,nopostdot,style=super,nonumberlist,toc,nogroupskip]{glossaries}

\glsdisablehyper 

\newcommand{\accolor}[1]{\textcolor{Black}{#1}}

\newacronymstyle{myacro}
{%
  \GlsUseAcrEntryDispStyle{long-short}%
}%
{%
  \GlsUseAcrStyleDefs{long-short}%
}

\setacronymstyle{myacro}

\newcommand*{\tip}[1]{
    \ifglsused{#1}{
      {\pdftooltip{\accolor{\glsentryshort{#1}}}{\glsentrydesc{#1}}}%
    }{%
      \gls{#1}
    }%
}%

\newcommand*{\tips}[1]{
    \ifglsused{#1}{
      {\pdftooltip{\accolor{\glsentryshortpl{#1}}}{\glsentrydescplural{#1}}}%
    }{%
      \glspl{#1}
    }%
}%

\newacronym{DVS}{DVS}{Dynamic Vision Sensor}
\newacronym{TC}{TC}{Temporal Contrast}
\newacronym{SNR}{SNR}{Signal-to-Noise Ratio}
\newacronym{SF}{SF}{Source-Follower}
\newacronym{theta_on}{$\Theta_{\text{ON}}$}{ON threshold}
\newacronym{theta_off}{$\Theta_{\text{OFF}}$}{OFF threshold}

\newacronym[description={False Negative Rate; signal that is incorrectly classified as noise}]{fnr}{FNR}{False Negative Rate}
\newacronym[description={False Negative; signal event that is incorrectly classified as noise event}]{fn}{FN}{False Negative}
\newacronym[description={False Positive Rate; noise that is incorrectly classified as signal}]{fpr}{FPR}{False Positive Rate}
\newacronym[description={False Positive; noise event that is incorrectly classified as signal event}]{fp}{FP}{False Positive}
\newacronym[description={Guided Event Filtering: Joint Filtering of Intensity Images and Neuromorphic Events}]{gef}{GEF}{Guided Event Flow}
\newacronym[description={Inter Spike Interval (nomenclature from neuroscience)}]{isi}{ISI}{Inter Spike Interval}
\newacronym[description={MAC (Multiply-Accumulate) is the basic operation of signal processing and artificial neural networks. One MAC is 2 Op.}]{mac}{MAC}{Multiply-Accumulate}
\newacronym[description={Multipurpose block random access memory module in FPGA}]{bram}{BRAM}{Block RAM}
\newacronym[description={Register Transfer Logic intermediate form, consisting of combinational and synchronous register logic cells}]{rtl}{RTL}{Register Transfer Logic}
\newacronym[description={Single Threshold Metric; a measure of the ROC TPR/FPR tradeoff at one discrimination threshold}]{stm}{STM}{Single Threshold Metric}
\newacronym[description={Surface of Active Event; image of latest event timestamps, same as Timestamp Image}]{sae}{SAE}{Surface of Active Events}
\newacronym[description={System on Chip; FPGA with embedded programmable processor}]{soc}{SoC}{System on Chip}
\newacronym[description={Time Surface; image of age of events relative to a particular event}]{ts}{TS}{Time Surface}
\newacronym[description={Timestamp Image; 2D image of latest event timstamps, similar to Surface of Active Events}]{ti}{TI}{Timestamp Image}
\newacronym[description={Timestamp+Polarity Image; 2D image of latest event timstamps and +/- brightness change polarites}]{tpi}{TPI}{Timestamp+Polarity Image}
\newacronym[description={True Negative Rate; noise that is correctly classified as noise}]{tnr}{TNR}{True Negative Rate}
\newacronym[description={True Negative; noise that is correctly classified as noise}]{tn}{TN}{True Negative}
\newacronym[description={True Positive Rate; signal that is correctly classified as signal}]{tpr}{TPR}{True Positive Rate}
\newacronym[description={True Positive; signal event that is correctly classified as signal}]{tp}{TP}{True Positive}
\newacronym[longplural={Convolutional Neural Networks}]{cnn}{CNN}{Convolutional Neural Network}
\newacronym[longplural={First In First Out memories}]{fifo}{FIFO}{First In First Out memory}
\newacronym{adc}{ADC}{Analog to Digital Converter}
\newacronym{aer}{AER}{Address Event Protocol}
\newacronym{aps}{APS}{Active Pixel Sensor}
\newacronym{asic}{ASIC}{Application Specific Integrated Circuit}
\newacronym{auc}{AUC}{Area Under the Curve}
\newacronym{baf}{BAF}{Background Activity Filter}
\newacronym{ba}{BA}{Background Activity}
\newacronym{bmof}{BMOF}{Block Matching Optical Flow}
\newacronym{bm}{BM}{Block Matching}
\newacronym{bp}{BP}{Back Propagation}
\newacronym{cfa}{CFA}{Color Filter Array}
\newacronym{cf}{CF}{Complementary Filter}
\newacronym{cg}{CG}{Convolutional Gated Recurrent Unit Network}
\newacronym{cis}{CIS}{CMOS Image Sensor}
\newacronym{cmae}{CMAE}{Cross-Modality Attention Enhancement}
\newacronym{contrastmaximization}{CM}{Contrast Maximization}
\newacronym{cots}{COTS}{Commodity Off-The-Shelf}
\newacronym{cpu}{CPU}{Central Processing Unit}
\newacronym{cv}{CV}{Computer Vision}
\newacronym{davis}{DAVIS}{Dynamic and Active pixel Vision Sensor}
\newacronym{dba}{DBA}{Dynamic Background Activity noise filtering algorithm}
\newacronym{dnn}{DNN}{Deep Neural Network}
\newacronym{dof}{DOF}{Degree of Freedom}
\newacronym{dolp}{DoLP}{Degree of Linear Polarization}
\newacronym{dram}{DRAM}{Dynamic RAM}
\newacronym{drcn}{DRCN}{Deep Recurrent Convolutional Network}
\newacronym{dr}{DR}{Dynamic Range}
\newacronym{dsp}{DSP}{Digital Signal Processing unit}
\newacronym{dvs}{DVS}{Dynamic Vision Sensor}
\newacronym{dwf}{DWF}{Double Window Filter}
\newacronym{e2pd}{E2PD}{Events to Polarization Dataset}
\newacronym{e2p}{E2P}{Events to Polarization}
\newacronym{edflow}{EDFLOW}{Event-driven Optical Flow}
\newacronym{edncnn}{EDnCNN}{Event Denoising CNN}
\newacronym{edp}{EDP}{Event Denoising Precision}
\newacronym{efast}{EFAST}{Event-Based time surface FAST}
\newacronym{epm}{EPM}{Event Probability Mask}
\newacronym{fast}{FAST}{Features from Accelerated Segment Test}
\newacronym{feast}{FEAST}{Feature Extraction with Adaptive Selection Thresholds }
\newacronym{flipflop}{FF}{Flip-Flop}
\newacronym{fom}{FOM}{Figure of Merit}
\newacronym{fpga}{FPGA}{Field Programmable Gate Array}
\newacronym{fpn}{FPN}{Fixed Pattern Noise}
\newacronym{fps}{FPS}{Frames Per Second}
\newacronym{fsae}{FSAE}{Filtered Surface of Active Events}
\newacronym{fwf}{FWF}{Fixed Window Filter}
\newacronym{gpu}{GPU}{Graphics Processing Unit}
\newacronym{gt}{GT}{Ground Truth}
\newacronym{hdl}{HDL}{Hardware Description Language}
\newacronym{hdr}{HDR}{high dynamic range}
\newacronym{hls}{HLS}{High Level Synthesis}
\newacronym{icm}{ICM}{Iterated Conditional Modes}
\newacronym{id}{ID}{Index Decay}
\newacronym{iir}{IIR}{Infinite Impulse Response}
\newacronym{imu}{IMU}{Inertial Measurement Unit}
\newacronym{inceptiveevent}{IE}{Inceptive Event}
\newacronym{iot}{IoT}{Internet of Things}
\newacronym{ip}{IP}{Intellectual Property}
\newacronym{its}{ITS}{Invariant Time Surface}
\newacronym{knn}{KNN}{$K$-Nearest-Neighbor clustering}
\newacronym{li}{LI}{Leaky Integrator}
\newacronym{lk}{LK}{Lucas-Kanade}
\newacronym{lpips}{LPIPS}{Learned Perceptual Image Patch Similarity}
\newacronym{lut}{LUT}{LookUp Table}
\newacronym{mlpf}{MLPF}{MultiLayer Perceptron denoising Filter}
\newacronym{mlp}{MLP}{Multilayer Perceptron}
\newacronym{ml}{ML}{Machine Learning}
\newacronym{mpeg}{MPEG}{Motion Picture Experts Group}
\newacronym{mse}{MSE}{Mean Squared Error}
\newacronym{na}{NA}{Numerical Aperture}
\newacronym{nnb}{NNb}{Nearest Neighbor}
\newacronym{of}{OF}{Optical Flow}
\newacronym{onf}{ONF}{Order(N) Filter}
\newacronym{pcb}{PCB}{Printed Circuit Board}
\newacronym{pdavis}{PDAVIS}{Polarization Dynamic and Active pixel VIsion Sensor}
\newacronym{pd}{PD}{photodiode}
\newacronym{pe}{PE}{Processing Element}
\newacronym{pfa}{PFA}{Polarization Filter Array}
\newacronym{pl}{PL}{programmable Logic}
\newacronym{por}{POR}{Positive Output Ratio}
\newacronym{prm}{PRM}{Pixel Rendering Module}
\newacronym{ps}{PS}{Processing System}
\newacronym{pugm}{PUGM}{Probabilistic Undirected Graph Model}
\newacronym{qwp}{QWP}{Quarter Wave Plate}
\newacronym{ram}{RAM}{Random Access Memory}
\newacronym{ransac}{RANSAC}{Random Sample and Consensus}
\newacronym{ratp}{RATP}{Recursive Adaptive Temporal Pooling}
\newacronym{rb}{RB}{Residual Block}
\newacronym{relu}{ReLU}{Rectified Linear Unit}
\newacronym{roc}{ROC}{Receiver Operating Characteristic}
\newacronym{roi}{ROI}{Region of Interest}
\newacronym{rpmd}{RPMD}{Relative Plausibility Measure of Denoising}
\newacronym{rpm}{RPM}{Revolutions per Minute}
\newacronym{rppp}{RPPP}{Rich Polarization Pattern Perception}
\newacronym{sad}{SAD}{Sum of Absolute Differences}
\newacronym{sm}{SM}{Supplementary Material}
\newacronym{snr}{SNR}{Signal to Noise Ratio}
\newacronym{soa}{SOA}{state of the art}
\newacronym{sram}{SRAM}{Static RAM}
\newacronym{stcf}{STCF}{SpatioTemporal Correlation Filter}
\newacronym{tda}{TDA}{Time Decay Adapted}
\newacronym{td}{TD}{Time Decay}
\newacronym{timsl}{TS}{time slice}
\newacronym{usb}{USB}{Universal Serial Bus}
\newacronym{vga}{VGA}{Video Graphics Adaptor}
\newacronym{vhdl}{VHDL}{Very High-Speed Integrated Circuit Hardware Description Language}
\newacronym{zoh}{ZOH}{Zero-Order Hold}

\def\cvprPaperID{43} 
\def\confName{CVPR}
\def\confYear{2023}

\newcommand{\vpd}{V_\text{pd}}
\newcommand{\ipr}{I_\text{pr}}
\newcommand{\ip}{I_\text{p}}
\newcommand{\vpr}{V_\text{pr}}
\newcommand{\isf}{I_\text{sf}}
\newcommand{\vsf}{V_\text{sf}}
\newcommand{\idiff}{I_\text{d}}
\newcommand{\vdiff}{V_\text{diff}}
\newcommand{\ion}{I_\text{on}}
\newcommand{\ioff}{I_\text{off}}
\newcommand{\irefr}{I_\text{refr}}

\newcommand{\trefr}{\Delta_\text{refr}}

\newcommand{\Ip}{I_\text{p}}

\begin{document}

\title{Shining light on the DVS pixel:\\A tutorial and discussion about biasing and optimization}

\author{Rui Graça, Brian McReynolds, Tobi Delbruck\\
Inst. of Neuroinformatics UZH/ETH Zurich,Switzerland\\
{\tt\small Corresponding author: rpgraca@ini.ethz.ch}
}
\maketitle

\begin{abstract}
   The operation of the \tip{DVS} event camera is controlled by the user through adjusting different bias parameters. These biases affect the response of the camera by controlling - among other parameters - the bandwidth, sensitivity, and maximum firing rate of the pixels. Besides determining the response of the camera to input signals, biases significantly impact its noise performance. Bias optimization is a multivariate process depending on the task and the scene, to which the user's knowledge about pixel design and non-idealities can be of great importance.

   In this paper, we go step-by-step along the signal pathway of the \tip{DVS} pixel, shining light on its low-level operation and  non-idealities, comparing pixel level measurements with array level measurements, and discussing how biasing and illumination affect the pixel's behavior. With the results and discussion presented, we aim to help \tip{DVS} users achieve more hardware-aware camera utilization and modelling.
   
\end{abstract}

\section{Introduction}
\label{sec:intro}

Over the past decade, neuromorphic or event-based \tips{DVS} \cite{lichtsteiner2008latencyasynchronous,finateu2020backilluminated,brandli2014latencyglobal,suh2020dynamicvision} gained significant attention as a disruptive sensing technology, demonstrating key performance advantages over conventional frame-based sensors, including wide intra-scene dynamic range, low latency and power consumption, and a data-sparse output capturing only the dynamic information in a visual scene \cite{gallego2022eventbased}. These benefits come at the cost of increased circuit complexity compared to frame-based sensors, which is both a blessing and a curse. On one hand, \tips{DVS} are highly adaptable, offering many degrees of freedom in performance through tunable biases; however, optimizing biases for a particular application is extremely difficult, even for experts in the technology. The goal if this paper is to present the challenging topic of bias optimization in an easy to understand way and provide specific guidelines for setting event camera biases based on task requirements and scene limitations.     

General biasing guidance is discussed in \cite{lichtsteiner2008latencyasynchronous, inivation_userguide, metavision_biases}, but descriptions are mostly qualitative and do not capture the trade-offs encountered when adjusting biases. Key performance metrics for DVS are measured and reported in \cite{Lichtsteiner2006-vt,Moeys2016-sk,Li2017-xq} for varied illumination levels; however, full characterization across varied bias settings is not reported.

In \cite{graca2021unravelingtheparadox,graca2023optimalbiasing}, we show how \tip{DVS} noise performance and bandwidth depend on illumination and photoreceptor bias, and  \cite{delbruck2021feedbackcontrol} presents a discussion regarding the effect of different biases on the output noise rate. In the same work, an algorithm based on feedback control that dynamically tunes the bias settings with the goal of keeping the output event rate within a programmable target window is presented. A discussion about how biasing affects the sharpness of the output of a event camera is presented in~\cite{dilmaghani2022controlandevaluation}.  In \cite{mcreynolds2022experimentalmethods}, we propose a \tip{DVS} characterization method to infer \tip{TC} event threshold, pixel bandwidth, and refractory period, and show and discuss how the inferred parameters depend on the biases.  

A good understanding of the pixel operation and the effect of the biases is also important for the development of realistic \tip{DVS} models and simulators~\cite{rebecq18esim,hu2021v2e,joubert2021eventcamera}. Accurate models are important to predict the response of a \tip{DVS} to an arbitrary scene, which can both help users optimize their set-up and generate of simulated event datasets from pre-existing frame-based datasets. In particular, \textsl{v2e}~\cite{hu2021v2e} explains and models \tip{DVS} motion blur and noise.

Most of the knowledge about \tip{DVS} operation existing in the literature, including that regarding the effect of each bias, is based on the empirical observation of the DVS output, often supported by a simplified model of the pixel operation. Therefore, many second-order effects and non-idealities are often overlooked. In this paper, we go a step further, and present and discuss how visual input signal is converted and encoded into a stream of ON and OFF events at the \tip{DVS} output, and how the different biases can be optimized according to the scene and the application requirements. Combined with looking at the relation between input and output, we present and discuss what is going on inside the \tip{DVS} pixel, supported by measurements of a test-pixel isolated from a DAVIS346 array.

During extensive measurements on internal pixel signals, we realized certain phenomena are far from intuitive and many aspects have generally been overlooked or oversimplified, especially regarding sensitivity and bandwidth. This oversimplification often leads to incorrect assumptions about how to bias or to model a \tip{DVS}. In this work, we present a thorough description of \tip{DVS} operation, including non-ideal second order effects, and how biasing affects operation. We present this description using terms that are simple enough for \tip{DVS} users with minimal circuit expertise yet still accurately describe the pixel's operation.

\subsection{Contributions and Outline}
\label{sec:contributions}

Our goal in this paper is to compile the most comprehensive description of the \tip{DVS} pixel available by explaining precisely how a signal produces events, and how pixel behavior is influenced by the multivariate combination of user-defined biases and physical non-idealities along the way. To accomplish this, \cref{sec:dvs_operation} reviews the the pixel response in general terms by presenting pixel-level node voltage measurements that illustrate a time varying signal propagating through each stage of a DVS pixel. \cref{sec:photoreceptor,sec:event_gen} describe the pixel stages and corresponding biases in detail and report key performance metrics and noise rates across a range of bias values. We include never-before-published node voltage measurements from a DAVIS346 pixel and array measurements to graphically communicate low-level circuit behaviors in an easy-to-digest format. In \cref{sec:multivariate_bias}, we aggregate this information to present a first-of-its-kind \href{https://docs.google.com/spreadsheets/d/1XaS3hkcjlbSG5gaMnlAy89rbsomILDgu/edit#gid=1310047800}{bias optimization spreadsheet tool} which recommends specific bias adjustments or "tweaks" across a comprehensive range of task requirements and scene limitations. This deep dive into pixel performance leads to novel interpretations of threshold levels drifting in time as a function of brightness, and that every event encodes information about signal, noise, and junction leakage, which have important implications for improving future vision algorithms.

\section{Basic DVS Operation}
\label{sec:dvs_operation}
\begin{figure}[t]
    \centering
    \includegraphics[width=\linewidth]{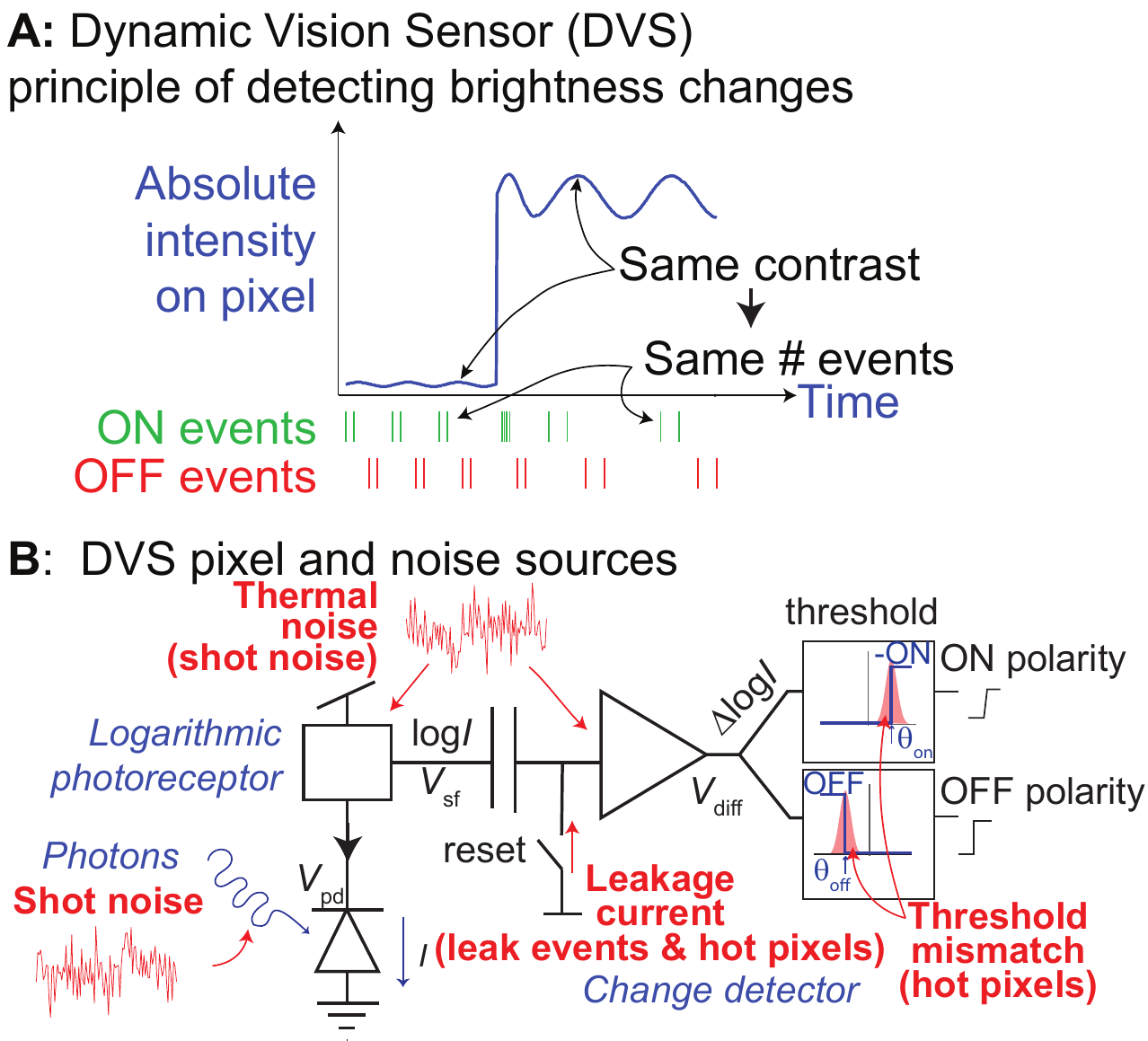}
    \caption{\textbf{A:} idealized output from \tip{dvs} pixel. \textbf{B:} Simplified DVS pixel circuit with main noise sources (adapted from \cite{guo2023lowcost}.)}
    \label{fig:dvsprinciple}
\end{figure}

The basic operation of the DVS pixel has been extensively covered in \cite{lichtsteiner2008latencyasynchronous,gallego2022eventbased,delbruck2021feedbackcontrol,hu2021v2e}. \cref{fig:dvsprinciple}A shows how each pixel converts an illumination signal changing in time to an output stream of events representing changes in relative light intensity.
\cref{fig:dvsprinciple}B shows a simplified pixel model. 
Input light converted to a voltage by a logarithmic photoreceptor, and changes in this voltage are encoded in events.
Events can be of \textit{ON} polarity, encoding an increase in light intensity, or of \textit{OFF} polarity, encoding a decrease in light intensity. An ON event is associated with \tip{theta_on} -- each ON event encodes that the input light intensity changed by a factor of $\exp(\Theta_\text{ON})$ since the last event (of either polarity). Similarly, each OFF event, associated with \tip{theta_off}, encodes a change by a factor of $\exp(-\Theta_\text{OFF})$ since the last event.
Noise and pixel-to-pixel mismatch affect the circuit through the sources illustrated.

\begin{figure*}[tb]
    \centering
    \includegraphics[width=\textwidth]{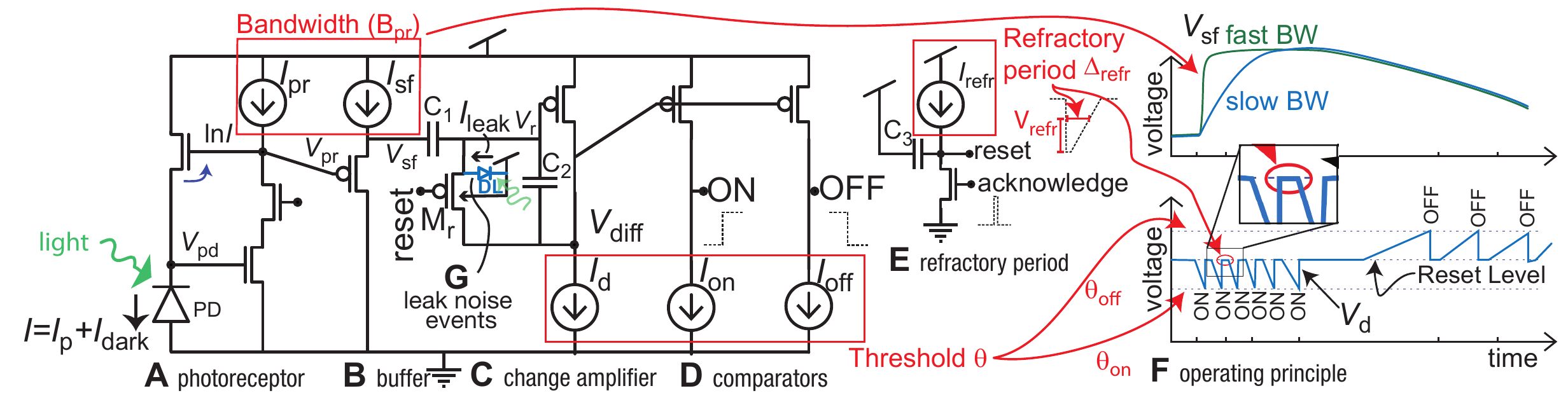}
    \caption{Typical DVS pixel circuit~\cite{taverni2018frontandback}. The active logarithmic photoreceptor (\textbf{A}) is buffered by a source-follower (\textbf{B}), which drives a cap-feedback change amplifier (\textbf{C}), which is reset on each event by a low-going \textit{reset} pulse.
    A finite refractory period (\textbf{E}) holds the change amplifier in reset for the refractory period $\trefr$. Comparators (\textbf{D}) detect ON and OFF events as seen in \textbf{F}. Periodic leak events result from junction and parasitic photocurrent $I_\text{leak}$ in diode \textsf{DL} (\textbf{G}). 
    }
    \label{fig:davis_circuit}
\end{figure*}

\cref{fig:davis_circuit} shows the detailed \tip{DVS} pixel circuit. 
The photoreceptor, \textbf{A}, consists of a photodiode \textbf{PD} that transduces light into an electric current $\Ip$, and a feedback loop that ensures that the output voltage signal $\vpr$ is proportional to the logarithm of $\ip$. Stage \textbf{B} cleans the signal by removing fast changes in $\vpr$, which is good when the signal of interest moves slower than the noise components. The maximum frequency of the signal allowed through this stage is called the pixel ``bandwidth,'' and depends on two biases: $\ipr$ and $\isf$, as well as the input illuminance. One important remark here is that the pixel bandwidth is a totally different concept from the camera's readout bandwidth, i.e., the total rate of events we can read out from the camera, which is limited by the readout circuitry and not by the pixel. 

Stage \textbf{C} amplifies voltage changes on node $\vsf$ by the ratio of the capacitors ($\frac{-C_1}{C_2}$), which is around 20 for the DAVIS346. The resulting amplified signal $\vdiff$ is then compared to an ON threshold proportional to $\log{(\ion/\idiff)}$ and to an OFF threshold proportional to $\log{(\idiff/\ioff)}$~\cite{nozaki2017temperatureandparasitic}. When either of the thresholds is met, either an ON or an OFF event is generated and transmitted to the outside of the array. At the same time, a reset signal is generated in \textbf{E}, that sets $\vdiff$ to a reset state value independent of the input and controlled uniquely by $\idiff$. $\vdiff$ is held at the reset state for a time period know as ``refractory period''. The duration of the refractory period is inversely proportional to its control bias $\irefr$, and during the refractory period the input signal is ignored by the pixel. This way, the pixel operation switches between a tracking mode, in which $\vdiff$ amplifies the input signal, and a reset mode after each event, where $\vdiff$ is reset to a known state and ignores input changes.

The equivalence between the bias names used in this paper (and generally adopted in the literature), their descriptive name, and the name used in iniVation documentation and software is presented in~\cref{tab:bias}.

\begin{table}[t]
    \centering
    \caption{Bias names used in this paper (\cref{fig:davis_circuit}) and their equivalent name for DAVIS346 and other iniVation cameras}
    \begin{tabular}{|l|l|l|}
    \hline
    Short name & DAVIS346 & Description \\ \hline
    $\ipr$ & PrBp & Photoreceptor bias \\ 
    $\isf$ & PrSFBp & Source Follower (Buffer) bias \\
    $\idiff$ & DiffBn & Change Amplifier bias \\
    $\ion$ & OnBn & ON Threshold bias \\
    $\ioff$ & OffBn & OFF Threshold bias \\
    $\irefr$ & RefrBp & Refractory bias \\
    \hline
    \end{tabular}
    \label{tab:bias}
\end{table}

\begin{figure}
    \centering
    \includegraphics[width=0.8\linewidth]{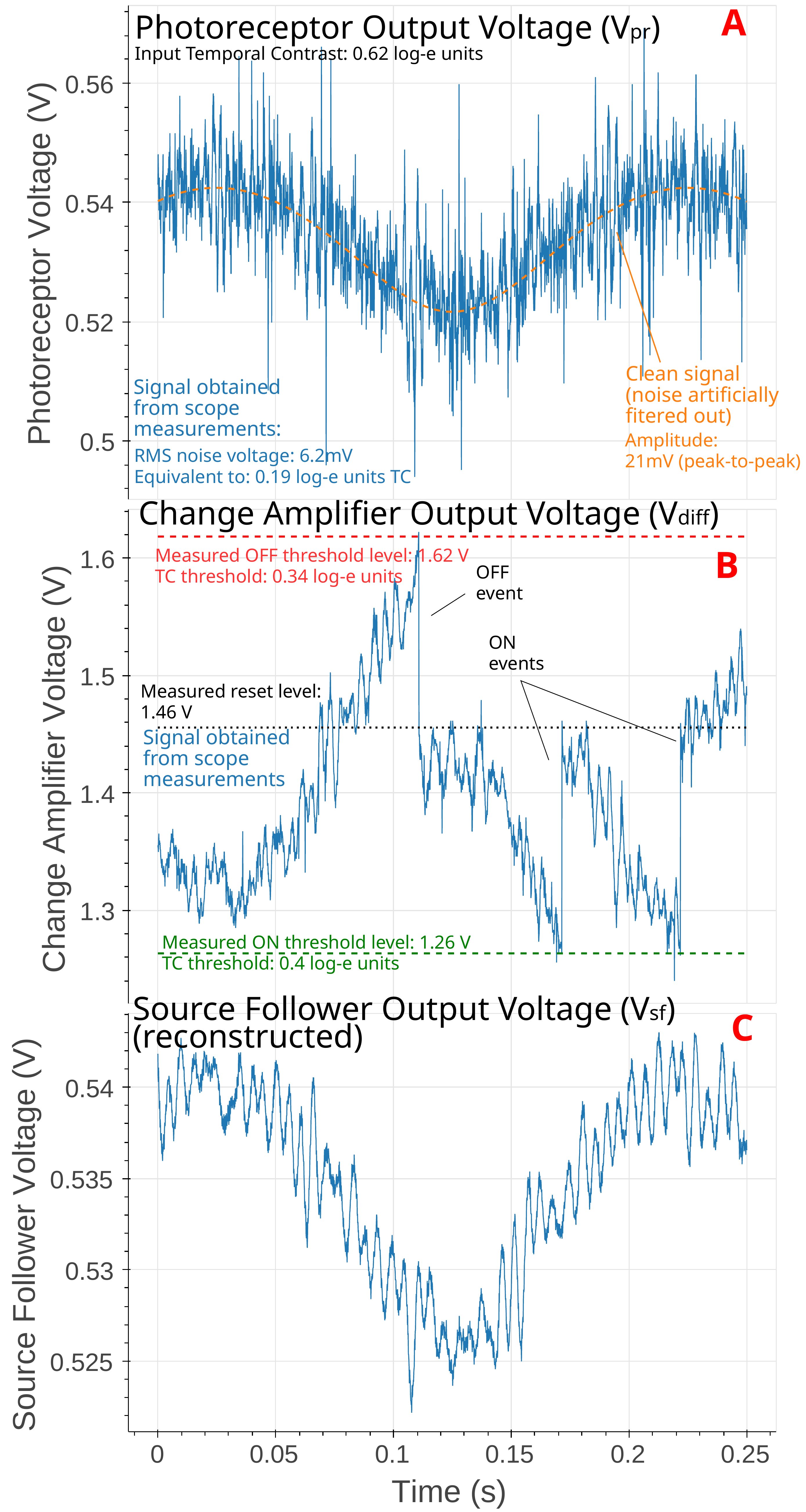}
    \caption{Test pixel recordings from internal nodes of the \tip{DVS} pixel in \cref{fig:davis_circuit}. \textbf{A} and \textbf{B} show the measured voltage at $\vpr$ and $\vdiff$ for a \SI{5}{\hertz} input sine wave with a \tip{TC} contrast of 0.62. \textbf{C} shows a computer reconstruction of $\vsf$ based in \textbf{A} and \textbf{B}, obtained by low-pass filtering \textbf{A} at a cutoff frequency of \SI{200}{\hertz}. $\ipr$, is biased strongly (large current), so the bandwidth is large and thus $\vpr$ is noisy. The fast frequency components are then filtered out by $\isf$}
    \label{fig:vpr_vdiff_vsf_scope}
\end{figure}

To illustrate pixel response at each stage, \cref{fig:vpr_vdiff_vsf_scope} presents direct measurements of node voltages $\vpr$ (\textbf{A}) and $\vdiff$ (\textbf{B}), and a computer-reconstructed $\vsf$ signal (\textbf{C}). The recording captures 1.25 periods of a \SI{5}{\hertz} sinusoidal input light signal with a contrast of 0.62 log-e units (meaning that the ratio between maximum and the minimum brightness of the sine wave is $e^{0.62}$ , or 1.86). This signal results in one OFF event (at t $\approx$0.12s) and two ON events (t $\approx$0.17s $\approx$0.22s). The measurements clearly demonstrate how noisy $\vpr$ is in practice, and how effectively bandwidth control with $\isf$ removes high frequency noise -- $\vsf$ is considerably cleaner, indicating a much more favorable \tip{SNR}. From $\vsf$ to $\vdiff$, the signal is inverted and amplified, resulting in a much larger peak-to-peak signal amplitude. Additionally, we can clearly see the effects of the reset logic after each threshold crossing, as $\vdiff$ returns to a fixed level immediately after each event. On this timescale (and bias configuration), the refractory period is too small to see. 

\section{The Logarithmic Photoreceptor}
\label{sec:photoreceptor}
\subsection{Phototransduction}
In the journey from incident photons to \tips{DVS} events, the first and crucial step is phototransduction. Ideally, every photon the lens captures is focused onto the photodiode and transfers its energy to break free an electron. The photoelectric effect dictates one photon generates a single "photoelectron," provided it has enough energy. Silicon is well suited to detect visible light, as only $\approx$1.1 eV of energy is needed to generate a photoelectron, and visible photons have 1.8 (red) - 3.1 (violet) eV.  These photoelectrons move freely when a voltage is applied, comprising "photocurrent".

In reality, several non-ideal behaviors influence photocurrent. Each photon has a finite probability of interacting with an atom in the silicon, and some photoelectrons will recombine with an atom when they encounter a vacancy. These factors decrease quantum efficiency -- essentially, the fraction of photons that become usable current. Additionally, even when there is no light on the pixel, the photodiode still conducts a small amount of \textit{dark current}, and photocurrent smaller than this value is indiscernible. 

Even when illumination is constant, photons and electrons do not arrive at a fixed rate. Instead, there is some statistical variability to the number of photons arriving on the pixel during a given time window, as well as in the number of electrons arriving at each node in the circuit. This fluctuation results in "shot noise", which is the dominant noise source in \tip{DVS} operation \cite{graca2021unravelingtheparadox,graca2023optimalbiasing} and the reason why the measured photoreceptor output in \cref{fig:vpr_vdiff_vsf_scope}\textbf{A} looks far noisier than the dashed orange signal obtained by averaging the photoreceptor output over many cycles.       

\subsection{Feedback logarithmic photoreceptor}
\begin{figure}
    \centering
    \includegraphics[width=0.45\textwidth]{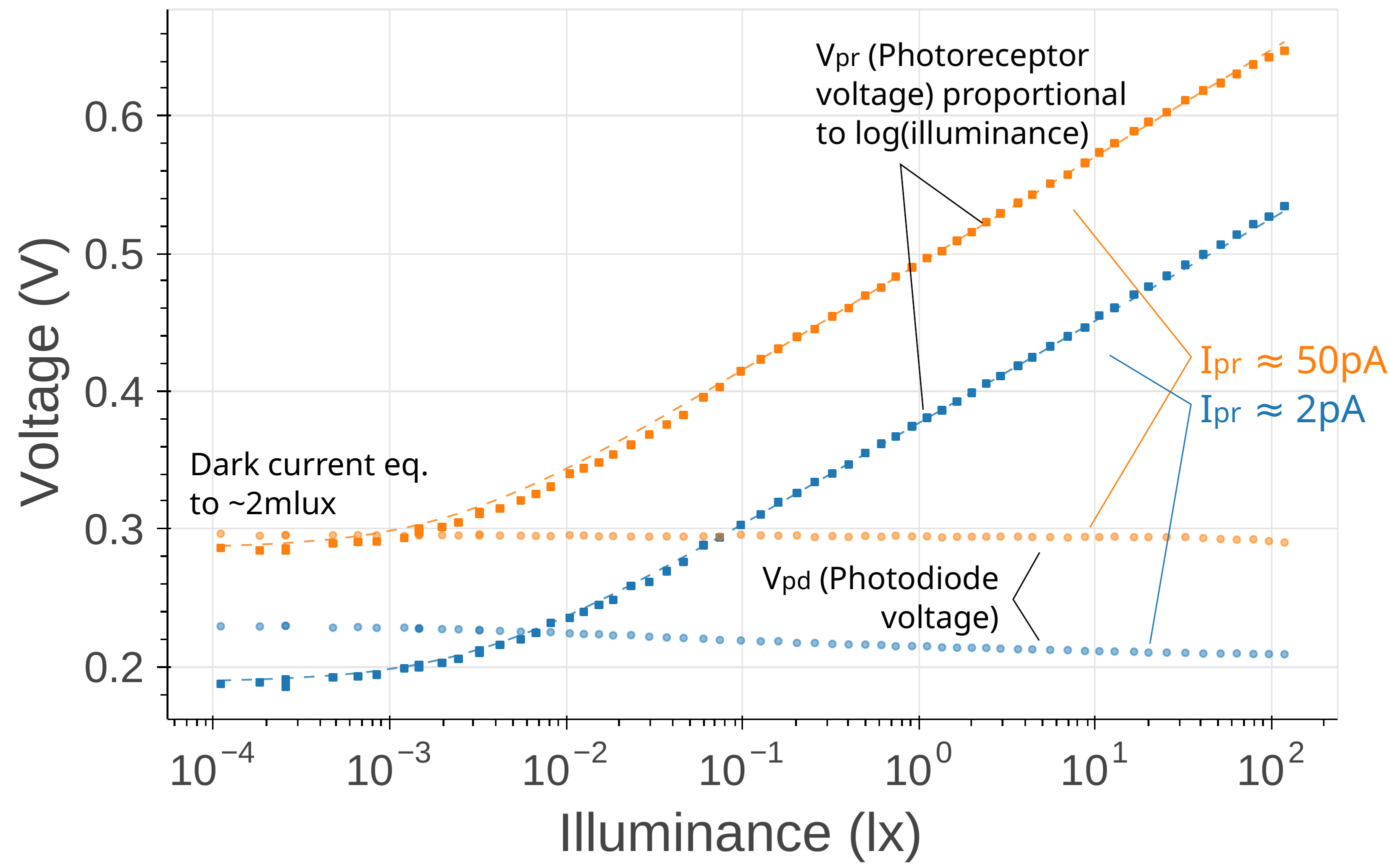}
    \caption{Photoreceptor voltage ($\vpr$) and Photodiode voltage ($\vpd$) versus illuminance for two different $\ipr$ settings. The dashed lines show a fit of a model of the photoreceptor including dark current: $\vpr=\frac{U_T}{\kappa}log(E_{v,\text{p}}+E_{v,\text{dark}})+V_{\text{p0}}$, where $E_{v,\text{p}}$ is the input illuminance, $E_{v,\text{dark}}$ is the illuminance equivalent to the dark current observed (fitted to approx. \SI{2}{\milli\lux}), $U_T$ is the thermal voltage ($\approx$\SI{25}{\milli\volt} at room temperature), $\kappa$ is a parameter of the transistor $\text{M}_{\text{fb}}$ and is fitted to approx. 0.75. $V_\text{p0}$ is a parameter determining the voltage level at \SI{1}{\lux}, and it varies with the logarithm of $I_{\text{pr}}$.}
    \label{fig:vpr_vs_illum}
\end{figure}

One of the key features of the \tip{dvs} is the logarithmic photoreceptor (\textbf{A} in \cref{fig:davis_circuit}) \cite{delbruck1995analog}. The logarithmic relationship between input light and output voltage enables two of the main advantages of the \tip{dvs}: first, it compresses the representation of the output signal, allowing a much higher dynamic range than conventional frame based cameras (which encode light intensity linearly); and second, it makes small changes around a background value proportional to the background value. This means that it directly encodes reflectance --- to first-order, the same object observed in different light settings will always result in the same output signal (\cref{fig:dvsprinciple}\textbf{A}). \cref{fig:vpr_vs_illum} shows how the photoreceptor voltages $\vpr$ and $\vpd$ change with input illuminance. $\vpd$ is nearly constant with illuminance at a value determined by $\ipr$, while $\vpr$ varies with the logarithm of input illuminance with a gain (slope) mostly independent of $\ipr$. The measurements also clearly show the lower limit of dynamic range, set by the dark current.

\subsection{Second-order effects of illumination}
\begin{figure}
    \centering
    \includegraphics[width=0.5\textwidth]{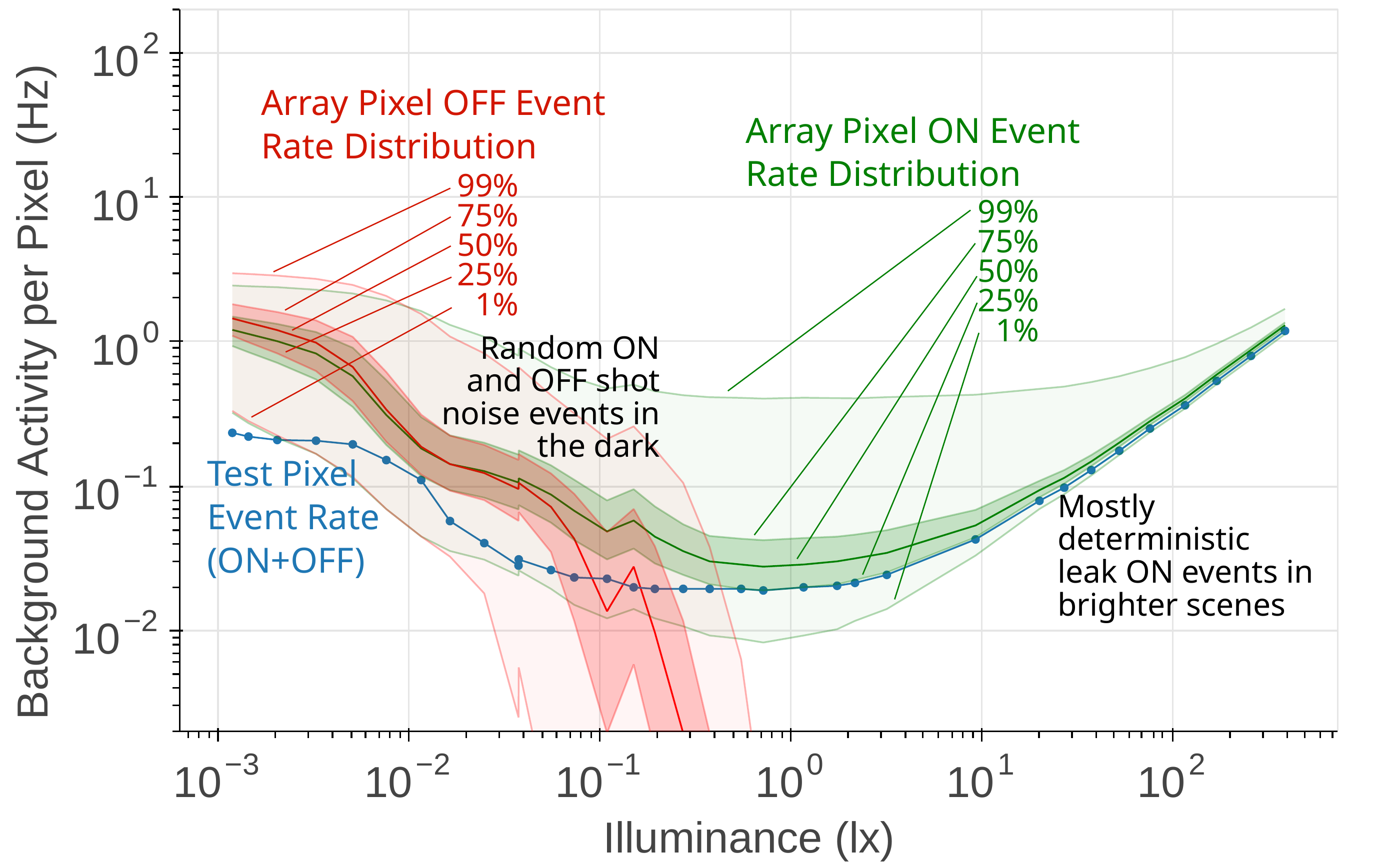}
    \caption{Distribution of the \tip{ba} event rate per pixel in a DAVIS346 biased with $\ipr$ of \SI{30}{\pico\ampere} and $\isf$ of \SI{15}{\pico\ampere} and nominal threshold and refractory period settings (tweaks of 0 in Figs. \ref{fig:tc_vs_threshTweak} and \ref{fig:refr_plots}) for varying on-chip illuminance. The green and red lines show the quantiles of the distribution of ON and OFF event rates in pixels of the array indicated in the figure. The blue line shows the event rate for a test pixel in the same camera, but isolated from the array. }
    \label{fig:noise_rate_vs_illum}
\end{figure}

Even though to a first order approximation the photoreceptor output is independent of the absolute illumination level, this is far from true for practical applications. Absolute illumination strongly affects noise and bandwidth, as discussed in~\cite{graca2021unravelingtheparadox}. The bandwidth of the photoreceptor is generally directly proportional to light intensity (unless $\ipr$ is set very low), meaning that in the dark, the ability to detect fast changes in the scene is limited and the latency from signal change to pixel response is longer. Also the \tip{ba} profile (i.e. the rate of "noise" events not encoding signal change) strongly depends on light intensity, as shown in \cref{fig:noise_rate_vs_illum}. The figure shows the \tip{ba} rate vs. illuminance for both the array and an isolated test pixel operating in the same conditions. As explained in~\cite{graca2021unravelingtheparadox,nozaki2017temperatureandparasitic}, the \tip{ba} consists of a high rate of random shot noise events of both polarities in the dark, exclusively ON events with rate proportional to light intensity in bright scenes (described in \cref{sec:event_gen}), and a dip in \tip{ba} between these two regions.  

\subsection{The Source-Follower Buffer}
The \tip{SF} Buffer (\textbf{B} in \cref{fig:davis_circuit}) serves two main purposes: it decouples the sensitive photoreceptor stage (\textbf{A}) from the spiking stages of the pixel (\textbf{C}, \textbf{D}, and \textbf{E}), and it can be used to limit the bandwidth, filtering out noise components faster than the signal of interest. This effect is visible in \cref{fig:vpr_vdiff_vsf_scope}, where $\vsf$ (\textbf{C}) and $\vdiff$ (\textbf{B}) exhibit significantly less noise (and of lower frequency components) than $\vpr$ (\textbf{A}). The bandwidth limit imposed by the \tip{SF} stage is proportional to its bias current $\isf$.

\subsection{Photoreceptor and Buffer biasing}
A discussion about optimal biasing of the photoreceptor and \tip{SF} buffer is presented in~\cite{graca2023optimalbiasing}. Both $\ipr$ and $\isf$ affect bandwidth and noise, but in different ways. The main practical difference is that $\ipr$ potentially introduces significantly more noise than $\isf$, but because it is downstream in the circuitry, $\isf$ can be used to remove excess noise introduced by $\ipr$. The higher $\ipr$, the faster the noise components it introduces, and the easier we can filter them out by adjusting $\isf$. As a result, contrary to what was proposed in \cite{delbruck2021feedbackcontrol}, the \tip{dvs} achieves significantly better noise performance when $\ipr$ is high and $\isf$ is low enough to filter out noise introduced by $\ipr$. $\isf$ can be used to limit signal bandwidth, but also to only remove high frequency noise components introduced by $\ipr$ without affecting signal bandwidth.

\begin{figure}
    \centering
    \includegraphics[width=0.5\textwidth]{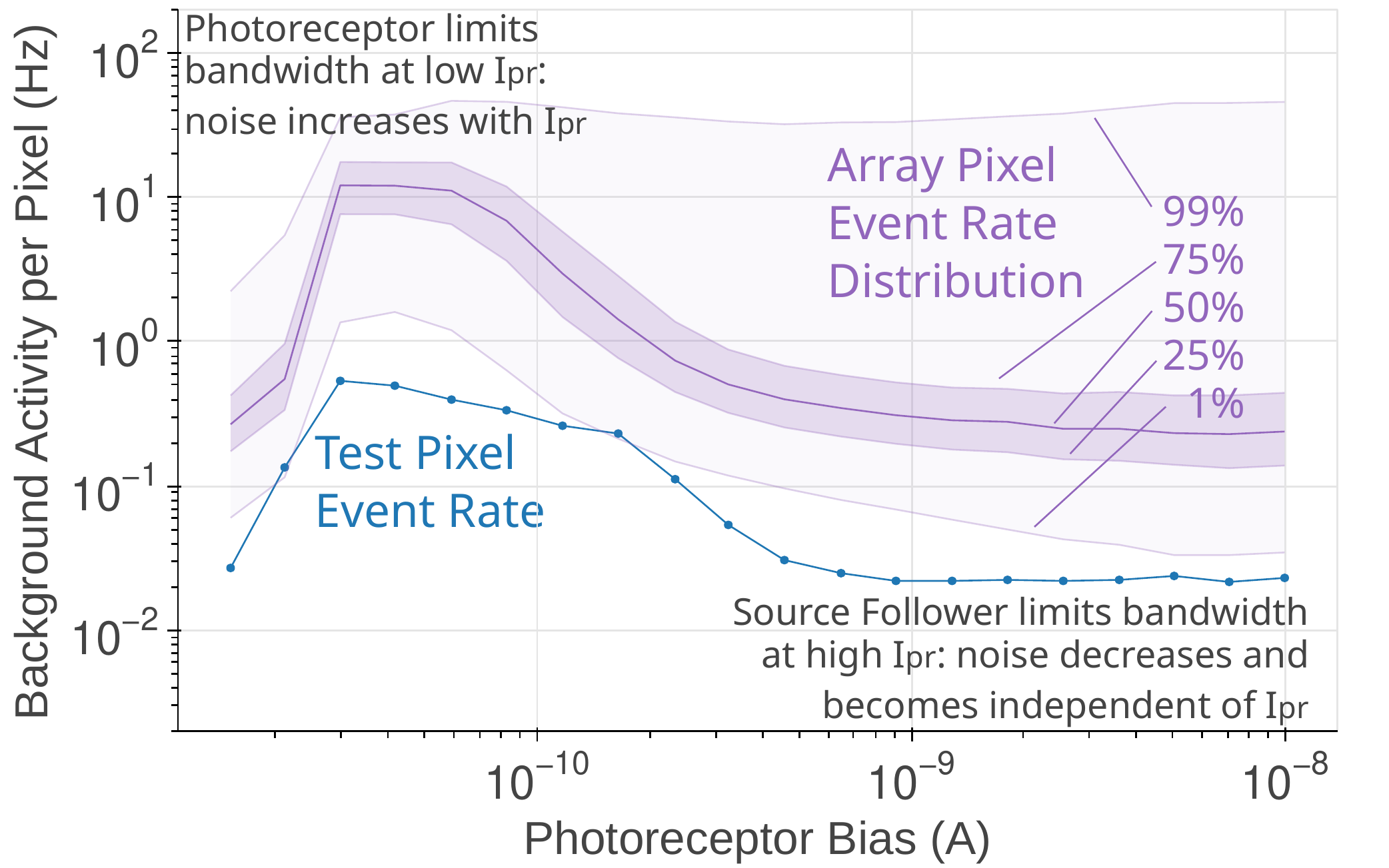}
    \caption{Distribution of the \tip{ba} per pixel in a DAVIS346 array (in purple) for varying $\ipr$ and for an isolated test pixel (in blue) at an on-chip illuminance of \SI{40}{\milli\lux} and $\isf$ of \SI{30}{\pico\ampere} and nominal threshold and refractory period settings}
    \label{fig:noise_rate_vs_PrBp}
\end{figure}

\cref{fig:noise_rate_vs_PrBp} shows the distribution of noise event rate per pixel in a \tip{dvs} camera in dark conditions (around \SI{40}{\milli\lux}) for varying $\ipr$. We see that for very low $\ipr$, noise increases with $\ipr$ (because bandwidth increases), but for higher $\ipr$ noise starts to decrease because the faster (higher frequency) noise components introduced by $\ipr$ are filtered out by the \tip{SF}. The results also show that from a certain point, increasing $\ipr$ has little effect on the output noise. This, combined with the fact that increasing $\ipr$ leads to higher power consumption, suggests that there is an optimal sweet spot for $\ipr$ (dependent on the selected $\isf$) at the lowest $\ipr$ where all of its noise components are filtered out. 

This point can be found empirically by finding the minimum $\ipr$ setting beyond which any increase will result in negligible noise rate reduction.
Under these circumstances, bandwidth is limited by (and proportional to) input light intensity up to an illumination level dependent on $\isf$, and then limited by $\isf$ at higher illumination levels. In~\cite{graca2023optimalbiasing}, we can see how bandwidth increases with $\isf$ , allowing higher frequency noise components to pass through -- leading to higher noise rates and pixel bandwidth.



\section{Event Generation}
\label{sec:event_gen}

\begin{figure*}[ht]
    \centering
    \includegraphics[width=0.9\textwidth]{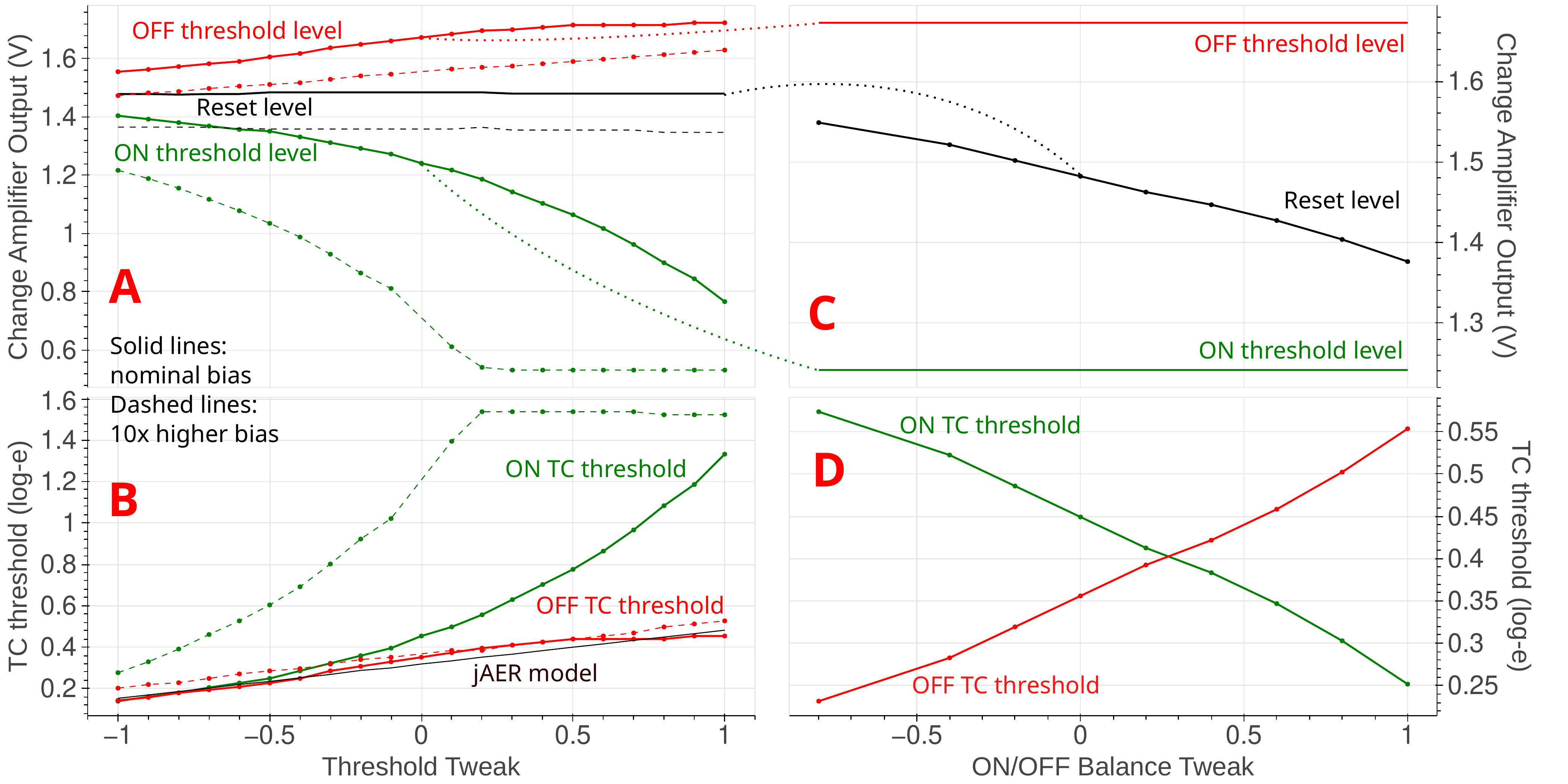}
    \caption{Threshold variation with jAER tweaks for threshold and ON/OFF balance. \textbf{A} shows the voltages measured at $\vdiff$ at which ON and OFF events are triggered (green and red lines) and to which the pixel is reset (in black). The solid lines were measured for nominal $\ion$, $\ioff$ and $\idiff$, and the dashed lines were measured for 10x higher $\ion$, $\ioff$ and $\idiff$. \textbf{B} shows the \tip{TC} threshold equivalent to the same measurements. \textbf{C} shows the same measured voltages as \textbf{A}, but for varying ON/OFF balance tweak, and \textbf{D} the equivalent \tip{TC} thresholds.}
    \label{fig:tc_vs_threshTweak}
\end{figure*}

The last step of the \tip{dvs} pixel is the conversion of a voltage signal to a stream of events, as described in \cref{sec:dvs_operation}. Understanding the process of event generation is fundamental to understanding how \tip{dvs} encodes data. One key aspect of event generation is that the state of the pixel (i.e. current illumination level) is stored (and reset after each event) in $\vdiff$ as an analog voltage. Unlike digital memory, analog memory destructively decays in time. In the case of the \tip{dvs}, this happens because of current that leaks through the reset switch (\textbf{G} in \cref{fig:davis_circuit}). The effect of leakage is that the voltage $\vdiff$ linearly decays in time at a rate that increases with absolute input illuminance. One way to look at this, is that $\vdiff$ is continually drifting toward an ON event, even in the absence of changes to the input signal. In this case, $\vdiff$ hits the ON threshold level at a fixed rate, generating so called "leak events," which dominate the \tip{ba} in brighter scenes, as depicted in~\cref{fig:noise_rate_vs_illum}~\cite{nozaki2017temperatureandparasitic}. An important note is that leakage is not noise -- its behavior is deterministic, and signal information is not destroyed but rather encoded in the timing between events.

Even though events are often discriminated into either ``signal'', ``noise'', or ``leak'' for simplification~\cite{hu2021v2e,guo2023lowcost,graca2021unravelingtheparadox}, it is important to note that an event never consists purely of one of these phenomena.  Rather, every event is a combination of all three. Noise causes not only the generation of events in the absence of signal changes, but also uncertainty (jitter) in the timing of an event in response to a signal change. Observing \cref{fig:vpr_vdiff_vsf_scope}, we can predict that events will most likely occur when noise adds constructively to the signal in the direction of the nearest threshold. In the same way, a small underthreshold signal can generate an event if noise adds constructively, and an over-threshold signal may not generate an event if noise adds destructively. Additionally, leakage makes it progressively harder to generate an OFF event (and easier to generate an ON event) as time increases after each reset. This can be interpreted as the OFF threshold increasing and the ON threshold decreasing linearly in time. These aspects of event generation influence how a \tip{DVS} should be biased, and should also be incorporated to design more robust event-processing algorithms. 

\subsection{Event Threshold}
The \tip{TC} event threshold quantifies how much the input signal needs to change since the last event to trigger the next event.   
An ON event is generated when $\vdiff$ drops low enough to toggle the output of the ON comparator (\textbf{D} in \cref{fig:davis_circuit}). This level is controlled by $\ion$. Likewise, $\ioff$ controls the level to which $\vdiff$ has to increase to generate an OFF event. This results in \tip{TC} thresholds, \tip{theta_on} and \tip{theta_off}, proportional to $\log{(\ion/\idiff)}$ and  $\log{(\idiff/\ioff)}$ respectively.

To study the effect of the \tip{TC} threshold, we adjusted the Threshold Tweak implemented in jAER described in~\cite{delbruck2021feedbackcontrol}. The Threshold Tweak is a user defined value between -1 and 1 that controls the \tip{TC} threshold by adjusting $\ion$ and $\ioff$. The solid green and red lines in \cref{fig:tc_vs_threshTweak}\textbf{A} show the measured $\vdiff$ values at which ON and an OFF events occur, and the solid black line shows the voltage to which $\vdiff$ is reset after each event. As expected, increasing the tweak increases the distance between the reset level and the threshold levels, but with two caveats: first, the ON threshold deviates from the desired linear variation with the tweak - this results from inaccurate modeling in jAER which can be corrected, and second, the OFF threshold saturates for higher tweaks - this happens because the value for $\ioff$ reaches the minimum allowed for proper circuit operation. The second condition can be alleviated by increasing all $\idiff$, $\ion$, and $\ioff$ by the same factor (at the expense of higher power consumption). The dashed lines show the corresponding levels when all currents are increased by a factor of 10. \cref{fig:tc_vs_threshTweak}\textbf{B} shows \tip{TC} thresholds equivalent to the same measurements, as well as the expected \tip{TC} level modeled by jAER. Considering the interpretation of leakage as a gradual decrease of \tip{theta_on} and increase of \tip{theta_off} following each reset, these \tip{TC} thresholds represent measured values immediately after reset.

\begin{figure}
    \centering
    \includegraphics[width=0.5\textwidth]{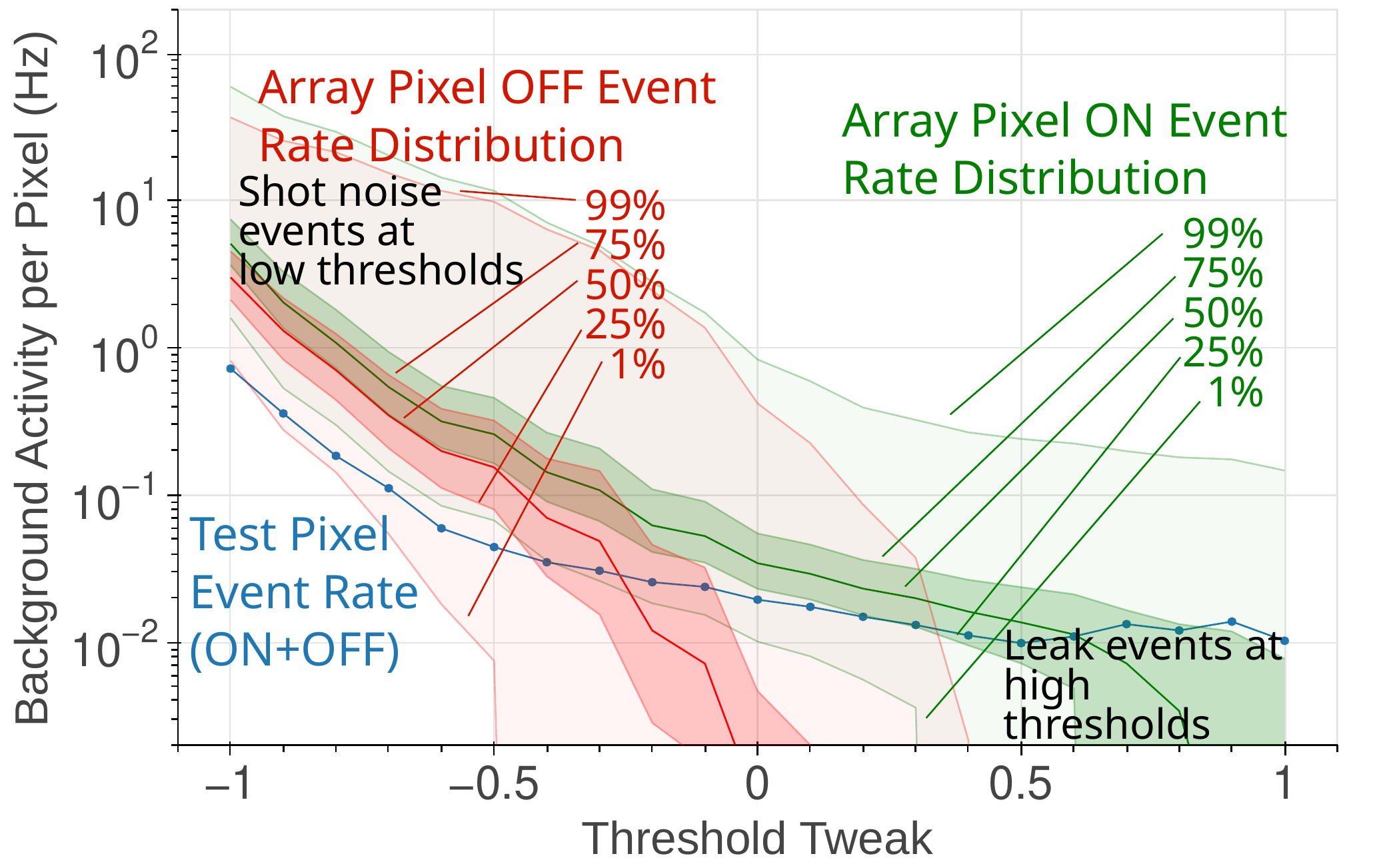}
    \caption{Distribution of ON (in green) and OFF (in red) \tip{ba} per pixel in the array and \tip{ba} rate in an isolated test pixel (in blue) for varying jAER Threshold Tweak at an on-chip illuminance of \SI{40}{\milli\lux}, $\ipr$ of \SI{3}{\nano\ampere}, $\isf$ of \SI{15}{\pico\ampere}, and nominal threshold and refractory period settings}
    \label{fig:noise_rate_vs_threshTweak}
\end{figure}

\cref{fig:noise_rate_vs_threshTweak} shows how \tip{ba} varies with the Threshold Tweak. As predicted by~\cite{graca2021unravelingtheparadox}, noise rates depend exponentially on the threshold. This dependence is given by the tail of the Gaussian function, since noise is approximately normally distributed. For higher thresholds, shot noise becomes negligible and leak events dominate the output.

\subsection{Balance between ON and OFF thresholds}
In ideal \tip{dvs} operation, for most applications, it would be desirable to have the same \tip{TC} threshold for ON and OFF events. However, as seen before, leakage introduces a fundamental imbalance between ON an OFF events, with a preference towards ON events. As discussed in~\cite{mcreynolds2023exploitingalternating}, deliberately applying imbalanced thresholds towards a more sensitive \tip{theta_on} can lead to better noise performance. 

The balance between \tip{theta_on} and \tip{theta_off} while keeping the same \tip{theta_on}$+$\tip{theta_off} can be adjusted by altering the bias $\idiff$, or equivalently, the ON/OFF balance tweak implemented in jAER, which adjusts $\idiff$ around a nominal value. The effect of the tweak is shown in \textbf{C} and \textbf{D} in \cref{fig:tc_vs_threshTweak}. In \textbf{C}, we can see that the voltage level at the output of the Change Amplifier that results in an ON or OFF event (green and red lines) is independent of the tweak, but the level to which $\vdiff$ resets after an event (black line) changes linearly with the tweak. In \textbf{D} we see how this translates to \tip{TC} threshold. 



\begin{figure}
    \centering
    \includegraphics[width=0.45\textwidth]{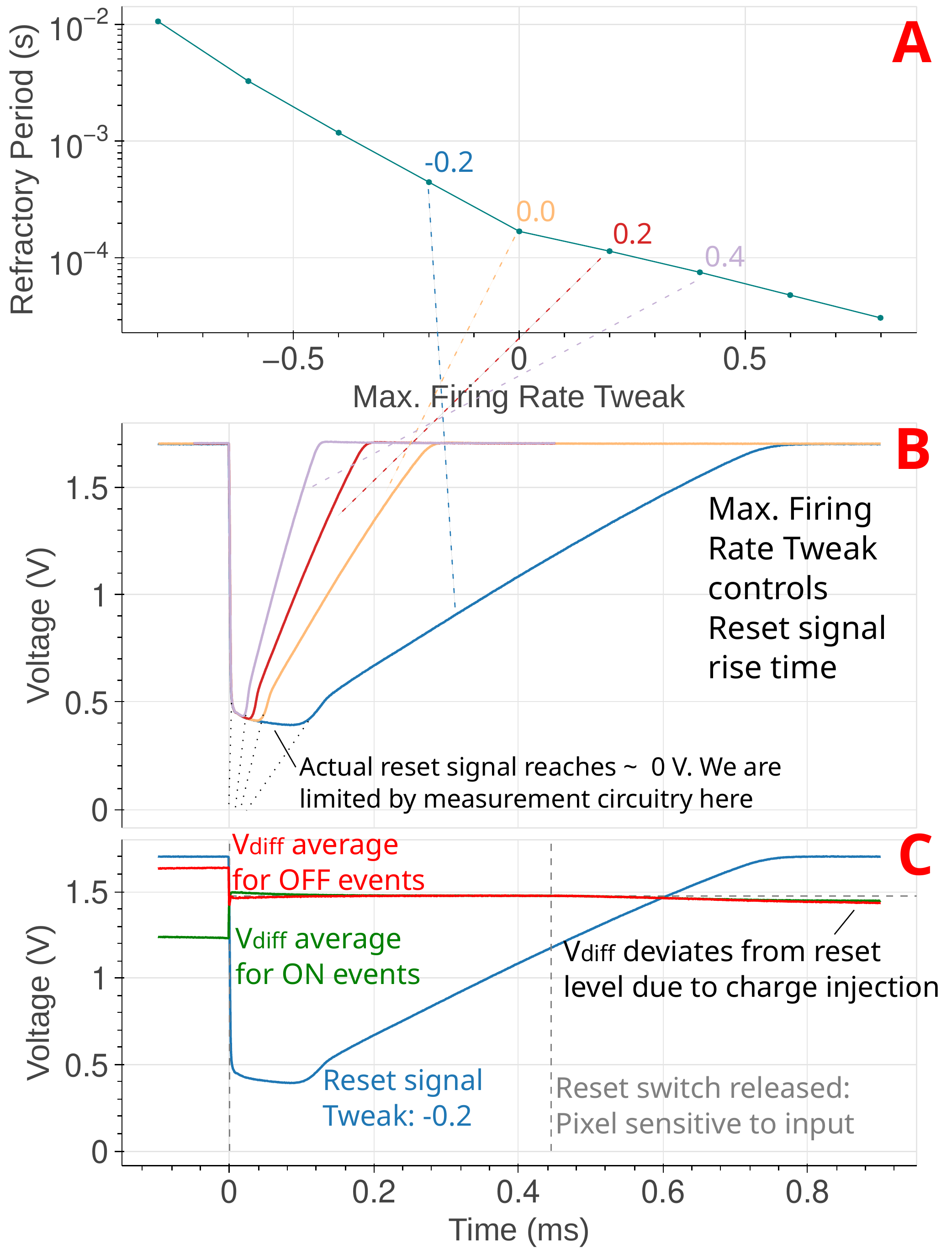}
    \caption{In-pixel measurements on the refractory period. \textbf{A} shows the length of the refractory period vs. the jAER Max Firing Rate Tweak. The length was obtained by directly measured the time during which the reset signal is active. \textbf{B} show average samples of measured pulses of the reset signal for different values of the tweak, with the correspondence with the points in \textbf{A} marked. \textbf{C} shows the same averaged reset pulse (for a tweak of \SI{-0.2}{}) and the average of $\vdiff$ for ON (in green) and OFF (in red) events. } 
    \label{fig:refr_plots}
\end{figure}
\begin{SCfigure*}[1][tb] 
    \centering    \includegraphics[width=.7\textwidth]{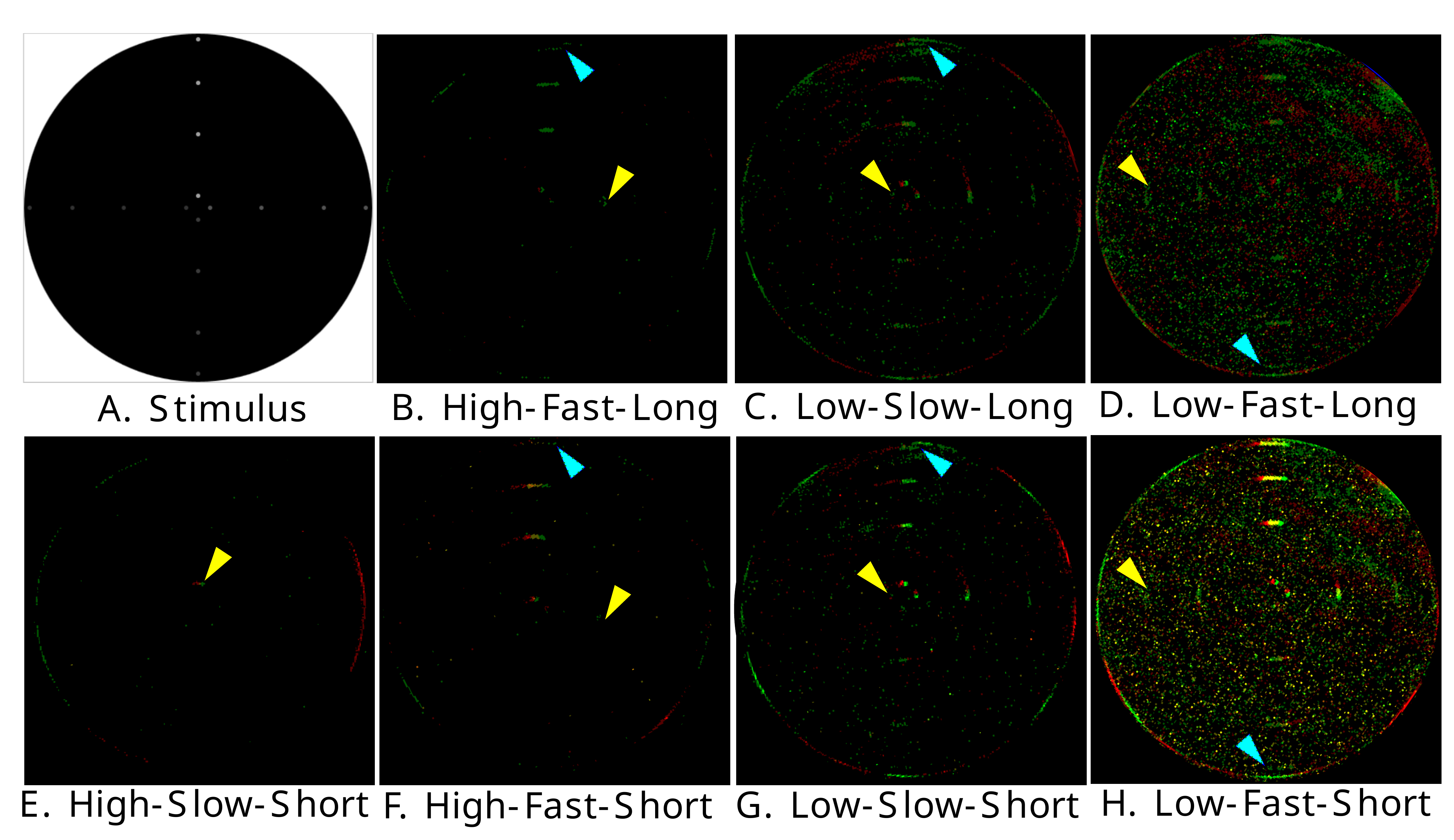}
    \caption{Multivariate biasing effects. \SI{10}{\milli\second} accumulations of \tip{DVS} events are shown while viewing a \SI{125}{RPM} motorized disk (\textbf{A}) with gray dots of varied contrast and speed (radial distance)
    under \SI{15}{\lux} illumination with a f/1.8 lens.  
    Events are green (ON) and red (OFF), and full scale rendering is 3 events.
    Each image (\textbf{B-H}) represents a combination of threshold (High/Low), bandwidth (Fast/Slow), and refractory period (Long/Short) biases. The leading edge of the dimmest (yellow) and fastest (blue) detectable dot is identified by an arrow.  If multiple dots of the same speed are detected, only the dimmest is indicated -- if multiple of the same brightness, only the fastest.}
    \label{fig:bias_effects}
\end{SCfigure*}

\subsection{Refractory period}
The duration of the refractory period is inversely proportional to its control bias $\irefr$. \cref{fig:refr_plots}\textbf{A} shows the measured length of the refractory period for varying values of the Max Firing Rate Tweak in jAER. A linear change in the tweak corresponds to an exponential change in $\irefr$. The refractory period was measured by directly observing the reset signal of a test pixel after an event and measuring the time the pixel is held in reset mode. Samples of pulses of the reset signal for different tweaks are shown in \cref{fig:refr_plots}\textbf{B}. The pixel is held in reset while the reset signal is low (below $\approx$ \SI{1.2}{\volt}). As shown, the length of the refractory period is controlled by changing the speed of the rising edge of the reset pulse. We observed that below a tweak of \SI{-0.8}{}, $\irefr$ is too weak to drive the pixel out of reset, and the pixel stops operating properly.
\cref{fig:refr_plots}\textbf{C} shows the measured voltage at $\vdiff$ following an ON (green) OFF (red) event, and the reset signal in blue. When the reset pulse drops, $\vdiff$ moves to the reset level ($\approx$\SI{1.5}{\volt}). When the reset signal nears this level, the pixel starts to respond to the input signal again. This is not depicted here since the input signal is not changing. However, right after the reset signal reaches \SI{1.2}{\volt}, $\vdiff$ drops slightly from the reset level. This effect is usually not problematic (it is equivalent to having a slightly lower reset level), but for refractory periods faster than around \SI{100}{\micro\second}, the amount of this deviation increases. For extremely short refractory periods, the operation of the pixel is severely affected - setting a practical upper limit to $\irefr$.

As discussed in \cite{delbruck2021feedbackcontrol,mcreynolds2023exploitingalternating}, $\irefr$ sets a trade-off between data fidelity and noise rate. A higher refractory period reduces noise, but of discards some signal information (e.g. consecutive events at an edge indicating a brighter object).


\section{Multivariate Bias Dependency}
\label{sec:multivariate_bias}

As \tip{DVS} are applied to increasingly diverse applications, it is important to consider both the low-level behaviors controlled by sensor biases and the multivariate effects of their combination. \cref{fig:bias_effects} provides a visualization of how varied bias combinations can have a drastic influence on sensor response to the same dynamic scene. The scene consists of a motorized disk with small white dots of different contrast and radial distance, rotating at \SI{125}{RPM} under dim indoor lighting conditions, using a scene illuminance of $\approx$ \SI{15}{\lux}.

From this small sample set, its clear that varied bias combinations result in drastically different performance. From left to right, the sensor becomes gradually more sensitive to faster and dimmer spots, but the noise also increases. In this scenario, the combination of bandwidth and threshold determines the fastest and dimmest detectable objects in a scene. The refractory period limits how much information is collected from each dot. For example, in \cref{fig:bias_effects}\textbf{D}, the long refractory period makes it more difficult to differentiate between objects of varied brightness compared to \textbf{H}, where the brightest dots clearly produce more events. All three parameters influence noise, and nuanced interactions can have a significant influence on overall noise rates, as demonstrated in \cite{mcreynolds2023exploitingalternating}.     

Although these visual examples are instructive, in a real application, there are more than just two sensing criteria to consider. To facilitate selecting an optimal bias configuration for widely varied scenarios, we created an online spreadsheet\footnote{CVPR 2023 - DVS Bias Optimization Tool:  {\tiny\url{https://docs.google.com/spreadsheets/d/1XaS3hkcjlbSG5gaMnlAy89rbsomILDgu/edit\#gid=1310047800}}} that recommends a combination of event camera biases based on six criteria that describe task requirements and scene limitations. The selected criteria are \textit{data priority} (i.e., whether higher event counts that include high-fidelity brightness information, or a data sparse representation of the visual field with lower event counts is preferred), \textit{sensor motion}, \textit{background illumination}, \textit{object size}, \textit{object contrast}, and \textit{object speed}. Each criteria is split into two categories, resulting in 64 categorical combinations, and for each we propose a set of bias "tweaks" to use as a starting point for optimizing event camera performance. Because different event camera models and software interfaces have slightly different bias formats, we describe bias tweaks in relative terms (i.e. fast-mid-slow bandwidth, low-mid-high sensitivity, long-mid-short refractory period).

\section{Conclusion}

The results and discussion presented in this paper bring light to the internal operation of the \tip{DVS} pixel, by showing how low-level effects such as noise and leakage affect signal, and by exploring how biasing can be optimized when these effects are considered. Moreover, we propose a novel way to interpret events as a combination of signal, noise, and leakage, and an interpenetration of leakage as a variation of \tip{TC} threshold in time. These approaches can lead to more optimal modeling and utilization of the \tip{DVS}. The proposed Bias Optimization Tool summarizes information described in the paper, and provides the \tip{DVS} user with an easy way to optimize biasing based on the task and scene, considering interaction of the main variables playing a role in the process.


\renewcommand*{\bibfont}{\footnotesize}
\printbibliography

\end{document}